\documentclass[conference]{IEEEtran}
\usepackage{amsthm}
\usepackage{amsmath}
\usepackage{amssymb}
\usepackage{subfigure}
\usepackage{epic}
\newtheorem{theorem}{Theorem}

\newtheorem{corollary}{Corollary}
\newcommand{\Nb}{{\mathbb{N}}}
\newcommand{\Zb}{{\mathbb{Z}}}
\newcommand{\Qb}{{\mathbb{Q}}}

\theoremstyle{remark}
\newtheorem{remark}{Remark}

\begin{document}

\title{Some limits to nonparametric estimation for ergodic processes}
\author{\IEEEauthorblockN{Hayato Takahashi}
\IEEEauthorblockA{The Institute of Statistical Mathematics,\\
10-3 Midori-cho, Tachikawa, Tokyo 190-8562, Japan.\\
 e-mail: hayato.takahashi@ieee.org.}}
 \markboth{}{}
 \maketitle
\begin{abstract}
A new negative result for nonparametric distribution estimation of binary ergodic processes is shown. 
The problem of estimation of distribution with any degree of accuracy is studied.
Then it is shown that  for any countable class of estimators there is a zero-entropy binary ergodic process that is inconsistent with the class of estimators.
Our result is  different from other negative results for universal forecasting scheme of ergodic processes. 
We also introduce a related result by B.~Weiss.
\end{abstract}
\begin{IEEEkeywords}
ergodic process, cutting and stacking, nonparametric estimation, computable function.   
\end{IEEEkeywords}
\IEEEpeerreviewmaketitle

\section{Introduction}\label{sec-intro}
Let \(X_1, X_2,\ldots\) be a binary-valued ergodic process and \(P\) be its distribution.
In this paper we study nonparametric estimation of binary-valued ergodic processes with any degree of accuracy. 
Let \(S\) and \(\Omega\) be the set of finite binary strings and the set of infinite binary sequences, respectively. 
Let \(\Delta(x):=\{x\omega | \omega\in\Omega\}\),
where \(x\omega\) is the concatenation of \(x\in S\) and \(\omega\), and write \(P(x)=P(\Delta(x))\).
For \(x\in S\), \(|x|\) is the length of \(x\).
Let  \(\Nb\), \(\Zb\), and \(\Qb\) be the set of natural numbers, the set of integers, and the set of rational numbers, respectively. 
From ergodic theorem, there is a function \(r\) such that  for \(x\in S\), \(n,k\in\Nb\),
\begin{equation}\label{rate-func}
\begin{gathered}
 P(\cup\{\Delta(y)\mid |P(x)- \frac{1}{|y|}\sum_{i=1}^{|y|-|x|+1} I _{y_i^{i+|x|-1}=x}|\geq 1/k, \\
\hspace{4.5cm} |y|=n\})<r(n,k,x), \\
 \forall x,k\ \lim_n r(n,k,x)=0,
\end{gathered}
\end{equation}
where   \(I\) is the indicator function and \(y_i^j=y_iy_{i+1}\cdots y_j\) for \(y=y_1\cdots y_n, i\leq j\leq n\).
\(r\) is called convergence rate.
If \(r\) is given, we know how much sample size is necessary to estimate the distribution with prescribed accuracy. 
However it is known that there is no universal convergence rate for ergodic theorem. 
If \(r\) is not known, ergodic theorem does not help to estimate the distribution with prescribed accuracy. 
Here a natural question arise: for any binary-valued ergodic process, is it always possible to estimate the distribution with any degree of accuracy with positive probability? 
We show that this problem has a negative answer, i.e.,  for any countable class of estimators there is a zero-entropy binary ergodic process that is not estimated from this class of estimators with positive probability. 
In particular, since the set of computable functions is countable, we see that there is a zero-entropy binary ergodic process that is inconsistent with computable estimators. 
Our result is not derived from other negative results for universal forecasting scheme of ergodic processes, see Remark~\ref{rem-forecasting}.

Let \(x\sqsubseteq y\) if \(x\) is a prefix of \(y\).
\(f\) is called estimator if 
\begin{equation}\label{f-condition}
\begin{aligned}
&f(x,k,y)\in\Qb\mbox{ is defined for } (x,k,y)\in S\times\Nb\times S\\
&\Rightarrow\forall z\sqsupseteq y\ f(x,k,z)=f(x,k,y).
\end{aligned}
\end{equation}
For \(\omega\in\Omega\), let \(f(x,k,\omega):=f(x,k,y)\) if \(f(x,k,y)\) is defined and \(y\sqsubset\omega\).
We say that \(f\) estimates \(P\) if
\begin{equation}\label{f-estimate}
\begin{aligned}
P(\omega \mid \forall x,k\ & f(x,k,\omega)\mbox{ is defined and }\\
&|P(x)-f(x,k,\omega)|<\frac{1}{k})>0.
\end{aligned}
\end{equation}
Here  \(\omega\) is a sample sequence and the minimum length of \(y\sqsubset\omega\) for which \(f(x,k,y)\) is defined is a stopping time.

In this paper, we construct an ergodic process that is not estimated from any given countable set of estimators:
\begin{theorem}\label{main-th-1}
\begin{align*}
&\forall F:\text{countable set of estimators~}\\
&\exists P\mbox{ ergodic and zero entropy }\forall f\in F\\
&P(\omega \mid \forall x,k\ f(x,k,\omega)\mbox{ is defined and }\\
&\hspace{2cm}|P(x)-f(x,k,\omega)|<\frac{1}{k})=0.
\end{align*}
\end{theorem}

We say that \(P\) is {\it effectively estimated} if there is a partial computable \(f\) that satisfies (\ref{f-condition}) and (\ref{f-estimate}).
Since the set of partial computable estimators is countable,  we have
\begin{corollary}\label{col-main-1}
There is a zero entropy ergodic process that is not effectively estimated. 
\end{corollary}

If \(r\) in (\ref{rate-func}) is computable then it is easy to see that  \(P\) is effectively estimated.
For example, i.i.d.~processes of finite alphabet are effectively estimated, see Leeuw et~al.~\cite{Leeuw56}.

As stated above, a difficulty of effective estimation of ergodic processes comes from that  there is no universal convergence rate for ergodic theorem. 
In Shields pp.171 \cite{shields96}, it is shown that for any given decreasing function \(r\), there is an ergodic process that satisfies  
\begin{equation}\label{counter-example}
\exists N\forall n\geq N\ P(|P(1)-\sum_{i=1}^{n}I_{X_i=1}/n|\geq1/2)>r(n).
\end{equation}
In particular if \(r\) is chosen such that \(r\) decreases to 0 asymptotically slower than any computable function 
then \(r\) is not computable. 
In V'yugin \cite{vyugin98}, a binary-valued computable stationary process with incomputable convergence rate is shown.

It is possible that an ergodic process is effectively estimated even if the convergence rate is not computable.
\begin{theorem}\label{main-th-2}
For any decreasing \(r\), there is a zero entropy ergodic process that is effectively estimated and satisfies (\ref{counter-example}).
\end{theorem}
For proofs of Thereom~\ref{main-th-1} and \ref{main-th-2}, see \cite{takahashi2011a}.
\begin{remark}
(i)  \(P\) is computable \(\Rightarrow\) (ii)  convergence rate \(r\) in (\ref{rate-func}) is upper semi-computable (effectively approximated from above) 
\(\Rightarrow\) (iii) \(P\) is effectively estimated. 
None of the converse is true. 
\end{remark}

\begin{remark}\label{rem-forecasting}
In Cover \cite{cover70}, two problems about prediction of ergodic processes are posed. 
Problem 1 : Is there a universal scheme \(f\) such that 
\(\lim_{n\to\infty} | f(X_0^{n-1})-P(X_n|X_0^{n-1})|\to 0\), a.s.  for all binary-valued ergodic \(P\)? Problem 2 : Is there a universal scheme \(f\) such that 
\(\lim_{n\to\infty} | f(X_n^{-1})-P(X_0|X_{-\infty}^{-1})|\to 0\), a.s.  for all binary-valued ergodic \(P\)?
Problem 2  was affirmatively solved by Ornstein \cite{{ornstein78},{weiss00}}.
Problem 1 has a negative answer as follows (Bailey, Ryabko, see \cite{{bailey},{ryabko88},{gyorfi-etal}}):
For any \(f\) there is a binary-valued ergodic process \(X_1, X_2,\ldots\) such that
\begin{equation}\label{bailey-ryabko}
P(\limsup_{n\to\infty} | f(X_0^{n-1})-P(X_n |X_0^{n-1})|>0)>0.
\end{equation}

It is not difficult to see that the above result is extended to a countable class \(\{f_1,f_2,\ldots\}\), i.e., for any \(\{f_1,f_2,\ldots\}\) there is an ergodic process such that    (\ref{bailey-ryabko}) holds for all \(f_1,f_2,\ldots\).
However this result does not imply Theorem~\ref{main-th-1}.
In fact, there is a finite-valued ergodic process that is effectively estimated but satisfies (\ref{bailey-ryabko}).
Roughly speaking, one of the difference between these problems is that in Problem 1 we have to estimate \(P(X_n |X_0^{n-1})\) from \(X_0^{n-1}\),
however in our estimation scheme, sample size is a stopping time and we can use a sufficiently large sample \(X_0^m, m>n\) to estimate \(P(X_0^n)\).
\end{remark}

\begin{remark}[B.~Weiss]
We say that \(f: S\to [0,1]\) is weakly universally consistent if  \(\forall\) binary ergodic \(P\) \(\forall\ \epsilon>0\ \exists\ N_{\epsilon}\ \forall n\geq N_{\epsilon}\)
\[
P( | P(1)-f(X_1^n) | <\epsilon )>1-\epsilon.\]
For example, \((X_1+\cdots+X_n)/n\) is weakly universally consistent. 
Then for any weakly universally consistent \(f\) and for any increasing \(n_1\leq  n_2\leq\cdots\) and \(\forall i\ \epsilon_i>0\)
there is a binary ergodic \(P\) and increasing sequence \(N_1\leq N_2\leq\cdots\) such that \(\forall i\ n_i\leq N_i\), \(P(1)=\frac{1}{2}\), and
\[
\forall k\geq 5\ P(| \frac{1}{2}-f(X_1^{N_k})|<\frac{1}{k})<\epsilon_k.
\]
In the above, it is easy to extend the sample size to a stopping time and \(f\) to a countable class of weakly universally consistent functions, i.e.,
for any countable class of weakly universally consistent functions \(\{f_1,f_2,\ldots\}\) and for any \(\epsilon_i>0, i=1,2,\ldots\), there is a binary ergodic \(P\) such that \(P(1)=\frac{1}{2}\), and
\begin{equation}\label{weiss}
\forall i\ \exists K\ \forall k\geq K\ P(\omega\mid | \frac{1}{2}-f_i(\omega)|<\frac{1}{k})<\epsilon_k.
\end{equation}
The difference between this result and Theorem~\ref{main-th-1} is that (\ref{weiss}) requires the universality of \(f\) but is  much stronger statement than  the fact that \(P\) is not estimated in 
the sense of Theorem~\ref{main-th-1}.
\end{remark}

\begin{center}
{\bf Acknowledgement}
\end{center}
The author thanks Prof.~Teturo Kamae (Matsuyama Univ.) and  Prof.~Benjamin Weiss (Hebrew Univ.) for helpful discussions and valuable comments.


\begin{thebibliography}{1}

\bibitem{bailey}
D.~H. Bailey.
\newblock {\em Sequential schemes for classifying and predicting ergodic
  processes}.
\newblock PhD thesis, Stanford Univ., 1976.

\bibitem{cover70}
T.~M. Cover.
\newblock Open problems in information theory.
\newblock In {\em 1975 IEEE Joint Workshop on Information Theory}, pages
  35--36, 1975.

\bibitem{Leeuw56}
K.~de~Leeuw, E.~F. Moore, C.~E. Shannon, and N.~Shapiro.
\newblock Computability by probabilistic machines.
\newblock In C.~E. Shannon and J.~McCarthy, editors, {\em Automata Studies},
  pages 183--212. Princeton Univ. Press, 1956.

\bibitem{gyorfi-etal}
L.~Gy{\"o}rfi, G.~Morvai, and S.~J. Yakowitz.
\newblock Limits to consistent on-line forcasting for ergodic time series.
\newblock {\em IEEE Trans.~Inform.~Theory}, 44(2):886--892, 1988.

\bibitem{ornstein78}
D.~S. Ornstein.
\newblock Guessing the next output of a stationary process.
\newblock {\em Israel J. Math.}, 30(3):292--296, 1978.

\bibitem{ryabko88}
B.~Ya. Ryabko.
\newblock Prediction of random sequences and universal coding.
\newblock {\em Probl.~Inf.~Transm.}, 24:87--96, 1988.

\bibitem{shields96}
P.~Shields.
\newblock {\em The ergodic theory of discrete sample paths}.
\newblock Amer.~Math.~Soc., 1996.

\bibitem{takahashi2011a}
H.~Takahashi
\newblock Computational limits to nonparametric estimation for ergodic processes.
\newblock http://arXiv.org/abs/1002.1559, submitted to IEEE trans IT.


\bibitem{vyugin98}
V.~V. V'yugin.
\newblock Ergodic theorems for individual random sequences.
\newblock {\em Theor.~Comp.~Sci.}, 207:343--361, 1998.

\bibitem{weiss00}
B.~Weiss.
\newblock {\em Single Orbit Dynamics}.
\newblock Amer.~Math.~Soc., 2000.

\end{thebibliography}
\end{document}